\documentclass[%
 aip,
 amsmath,amssymb,
preprint,
 reprint,%
]{revtex4-1}

\usepackage{graphicx}% Include figure files
\usepackage{dcolumn}% Align table columns on decimal point
\usepackage{bm}% bold math
\usepackage[utf8]{inputenc}
\usepackage[T1]{fontenc}
\usepackage{mathptmx}
\usepackage{etoolbox}
\usepackage{xcolor}
\usepackage{verbatim}

\makeatletter
\def\@email#1#2{%
 \endgroup
 \patchcmd{\titleblock@produce}
  {\frontmatter@RRAPformat}
  {\frontmatter@RRAPformat{\produce@RRAP{*#1\href{mailto:#2}{#2}}}\frontmatter@RRAPformat}
  {}{}
}%
\makeatother
\begin{document}

\preprint{}

\title[Stabilization of beam heated plasmas by beam modulation]{Stabilization of beam heated plasmas by beam modulation}
% Force line breaks with \\
\author{L. Einkemmer}
 \email{lukas.einkemmer@uibk.ac.at}
\affiliation{ 
Unversit\"at Innsbruck, \textcolor{black}{A-6020 Innsbruck,} Austria%\\This line break forced with \textbackslash\textbackslash
}%

\date{\today}% It is always \today, today,
             %  but any date may be explicitly specified

\begin{abstract}
A constant intensity beam that propagates into a stationary plasma
results in a bump-on-tail feature in velocity space. This results
in an instability that transfers kinetic energy from the plasma to
the electric field. We show that there are intensity profiles for
the beam (found by numerical optimization) that can largely suppress
this instability and drive the system into a state that, after the
beam has been switched off, remains stable over long times. The modulated
beam intensity requires no feedback, i.e.~no knowledge of the physical
system during time evolution is required, and the frequency of the
modulation scales approximately inversely with system size, which
is particularly favorable for large plasma systems. We also show that
the results obtained are robust in the sense that perturbations, e.g.~deviation
from the optimized beam profiles, can be tolerated without losing
the ability to suppress the instability.
\end{abstract}

\maketitle

\section{Introduction}

A particle beam that propagates into a stationary plasma can drive
a microinstability. Due to the additional peak (the beam) in the velocity
distribution that is superimposed on the Maxwellian equilibrium, this
is called the bump-on-tail instability. The instability is driven
by the unfavorable slope of the velocity distribution, which allows
the transfer of energy from fast particles to the electric field.
The electric field then grows exponentially until nonlinear effects
become strong enough to lead to saturation.

The bump-on-tail instability has been studied extensively (see, e.g.,
\cite{Denavit1985,Yamagiwa1987,Carlevaro2015,Zeeland2021}). It is
relevant for fusion plasmas as beam heating (the injection of high
energy particles to heat a plasma; see, e.g., \cite{Speth1989}) and
ion cyclotron heating (the heating of the plasma by absorption of
electromagnetic radiation) creates bump-on-tail features (see, e.g.,
\cite{Yamagiwa1987}). Bump-on-tail features from plasma heating as
well as runaway electrons can destabilize Alfvén waves \cite{Breizman2019}
and potentially degrade reactor performance \cite{Liu2021,Zeeland2021}.
For example, in \cite{Zeeland2021} it has been demonstrated that
simple beam modulation (turning the beam on and off with a certain
frequency) can significantly effect the ion transport even if the
same amount of heating is injected into the plasma. It has also been
pointed out \cite{Jing2024} that the bump-on-tail instability plays
a significant role in plasma thrusters (such as in the variable specific
impulse magnetoplasma rocket -- VASIMR).

A natural question then is whether a sufficiently strong beam necessarily
leads to an instability or if some type of control can be achieved.
Control theory is well established, but such schemes usually require
feedback that is obtained by taking measurements from the system under
investigation. Based on these measurements certain actions are then
taken to stabilize the system or achieve some other goal \cite{Burger2020,Albi2024}.
The classic example is the inverted pendulum stabilized by a PID controller.
However, feedback, poses a severe constraint for kinetic instabilities
that often take place on very fast time scales (say the inverse of
the plasma frequency or the Alfvén time). Thus, such an approach would
require that data acquisition, data processing, and the subsequent
modification of the control variable (say the intensity of the beam)
is fast enough to follow the plasma dynamics. In most situations this
is not feasible. In addition, beam-plasma instabilities are inherently
driven by non-Maxwellian velocity features, for which in most cases
only limited information can be obtained experimentally.

While feedback based control might be infeasible for microinstabilities,
it has been observed that modulating the beam intensity with a fixed
frequency can reduce the severity of the bump-on-tail instability.
In \cite{Nakamura1970} it was experimentally demonstrated that modulating
an electron beam close to the natural oscillation frequency of the
system reduces the electric field strength observed during the instability.
Later this problem was also studied by computer simulations. In \cite{Fukumasa1982}
the authors considered the nonlinear coupling of a relatively small
number of modes finding similar results. In \cite{Qin2014} a related
problem is investigated where the velocity (not the intensity) of
the beam is modulated. Neglecting thermal effects, the authors develop
a linear theory that shows that varying the velocity of the beam sinusoidally
reduces the growth rate of the instability (the frequency of modulation
is close to the plasma frequency). Similar observation, i.e.~that
a sinusoidal modulation can have a stabilizing effect, have also been
made in the context of inertial confinement fusion, where it is desirable
to suppress or delay a Rayleigh--Taylor instability. The interpretation
given in \cite{Kawata2018} is that an instability can be counteracted
if the phase is known. While measuring the phase of a naturally developing
instability is difficult in practice, a modulated beam can define
the phase of the perturbation and thus the instability can be counteracted.

The problem of suppressing a two-stream/bump-on-tail instability with
a time and space dependent external electric field as control has
also been considered recently \cite{Einkemmer2024a,Einkemmer2024b}.
The theoretical results obtained indicate that the instability can
be completely suppressed without feedback if the initial state is
known. The issue, however, is that injecting an arbitrary electric
field into a plasma is difficult in practice and that the initial
deviation from equilibrium needs to be precisely known. Nevertheless,
it raises the question what the ultimate limit of control is in situations
where the control is less direct, such as the case of beam modulation
that we consider here. 

In this paper our goal is to study a simple kinetic model of a stationary
plasma in which a (possibly modulated) beam is injected. The question
is whether the associated bump-on-tail instability can be suppressed \textcolor{black}{or reduced}
without requiring feedback and, if so, how fast this modulation needs
to be in order to be successful. Our goal is also to find beam profiles
by numerical optimization that are more effective than modulating
the beam in a sinusoidal manner.

\section{Model\protect\label{sec:model}}

For the electron dynamics we consider the one dimensional non-dimensionalized
Vlasov--Poisson equation

\begin{equation}
\partial_{t}f+v\partial_{x}f-E\partial_{v}f=S,\label{eq:vlasov-with-source}
\end{equation}
where $f(t,x,v)$ is the electron density and the beam is modeled
by a source term
\begin{equation}
\begin{aligned}
S(t,x,v)&=I(t)\frac{1}{\sqrt{2\pi}\sigma}\exp\left(\frac{-(x-x_{0})^{2}}{2\sigma^{2}}\right) \\
  &\qquad\qquad  \cdot\frac{1}{\sqrt{2\pi}v_{th,b}}\exp\left(\frac{-(v-v_{b})^{2}}{2v_{th,b}^{2}}\right),
  \end{aligned}\label{eq:source}
\end{equation}
where $I(t)$ is the modulated beam amplitude. We assume that the
beam has a Gaussian profile in space (centered at the middle of the
domain $x_{0}$; the spatial extent of the beam is $\sigma$) and
is Maxwellian in velocity (with average beam velocity $v_{b}$ and
thermal velocity \textcolor{black}{of the beam} $v_{th,b}$). We assume that the ions form a homogeneous
neutralizing background. Thus the electric field is given by $E=-\partial_{x}\phi$
with the potential $\phi$ satisfying $-\partial_{xx}\phi=\rho$.
The charge density $\rho$ is given by $\rho=\int f\,d(x,v)-\int f\,dv$.
We use a domain of size $L$ and periodic boundary conditions. Note
that as the beam adds particles to the system the number density is
time dependent and is given by $\int f(t,x,v)\,d(x,v)=1+\int_{0}^{t}I(s)\,ds$.
In principle we could also modulate the spatial extent $\sigma$,
the beam velocity $v_{b}$, and the thermal speed $v_{th,b}$, but
to keep the control relatively straightforward we will not do so here.
All units of time and velocity are normalized to the plasma frequency
($\omega_{p}$) and thermal velocity ($v_{th}$) of the bulk plasma \textcolor{black}{(i.e.~the thermal velocity of the initial electron distribution)},
respectively. This implies units of Debye length ($\lambda_{D})$
for all spatial variables. The electric field is normalized such that
no additional factor appears in the Poisson equation for the potential.
More specifically, in SI units we have $E=E_{\text{SI}}/E_{\text{ref}}$
with $E_{\text{ref}}=m\lambda_{D}\omega_{p}^{2}/e$, where $m$  is the electron mass and and
$e$ the unit charge. 

\begin{figure*}
\begin{centering}
\includegraphics[width=\textwidth]{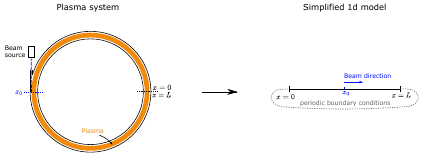}
\end{centering}
\caption{\textcolor{black}{We consider a plasma system into which a beam is injected. The simplified 1d model used in this study (shown on the right) consists of a domain of length $L$ with periodic boundary conditions. The beam is injected at position $x_0$ with velocity $v_b > 0$. This can be considered as an idealized version of a toroidal plasma system (as shown on the left) or alternatively a setup similar to \cite{Nakamura1970}, where a beam is reflected at the end of a vessel that encloses the plasma.} \label{fig:illustration}}
\end{figure*}

The model outlined is a fundamental model in plasma physics that has
been studied extensively both using linear theory and (usually nonlinear)
computer simulations. It is well known that adding a second peak (the
bump on the tail) to a Maxwellian velocity distribution, above a certain
threshold intensity, results in a plasma instability. Theoretically
this can be analyzed either by considering a separate fluid description
of the bulk plasma and the beam \cite{Chen2016} or, more accurately,
by deriving the dispersion relation within a fully kinetic description
\cite{Swanson2020}. In the latter case, the Penrose criterion \cite{Penrose1960}
is not satisfied and the bump-on-tail configuration is unstable. The
associated dispersion relation can be solved to obtain the (linear)
growth rate of the instability. The linear regime is accompanied by
an exponential increase of the electric energy, i.e.~kinetic energy
of the plasma is transferred to the electric field. Eventually the
system becomes strongly nonlinear and the amplitude of the electric
field saturates; the latter is not captured by linear theory. The described analysis is concerned with a static situation. That is, it is
assumed that, in some way, the system is prepared in a state where
a bump-on-tail feature is present. The plasma system is then allowed
to evolve on its own without further intervention. Here, in contrast,
we are, interested in a dynamic situation. That is, the plasma is
initially \textcolor{black}{a current free and spatially homogeneous} Maxwellian and the beam, modeled by the source term $S$,
dynamically creates the bump-on-tail feature and triggers the instability. \textcolor{black}{This setup is illustrated in Figure \ref{fig:illustration}.}

In our model of the beam we have four parameters: the beam amplitude
$I$, the spatial extent $\sigma$, the beam velocity $v_b$, and the beam thermal velocity
$v_{th,b}$. Note that the thermal velocity has a significant influence.
Increasing the thermal velocity such that it is comparable to $v_{b}$
softens the bump-on-tail feature and makes the system more stable.
Thus, we will restrict $v_{th,b}$ to lie between $0.2$ and $0.6$.
The position at which the beam enters the plasma $x_{0}$ in our model
just corresponds to a shift of the $x$ coordinates and is thus not
relevant. The beam velocity $v_{b}$ determines how much energy is
added to the system (for a given beam strength). We thus fix the total
energy that is added to the system by mandating that $v_{b}=3.5$
and $\int_{0}^{t_{b}}I(t)\,dt=I_{0}t_{b}$, where $I_{0}=0.05$ and
$t_{b}$ is the duration during which the beam is switched on. 
The goal is then to find $I(t)$, $\sigma$, and $v_{th,b}$ such
that the bump-on-tail instability is suppressed. 

\section{Results for sinusoidal beam profiles\protect\label{sec:sinusoidal}}

First we consider the case of a sinusoidal beam profile $I(t)=I_{0}+b_{1}\sin(\omega t)$, \textcolor{black}{$I_0=0.05$},
for $t\leq t_{b}$ and fixed beam temperature
($v_{th,b}=0.5$). We consider three configurations that correspond
to three different plasma sizes. Specifically, we choose the length
of the spatial domain as $L=4\pi$, $L=16\pi$, and $L=64\pi$. The
rational here is that ultimately we are interested in plasmas systems
that are large compared to the Debye length. Thus, we want to study
how the control schemes (especially the driving frequency, \textcolor{black}{but also the beam width}) scales
with systems size, as this dictates how fast the beam has to be modulated and \textcolor{black}{how tightly it needs to be focused}
to achieve the desired effect. \textcolor{black}{The beam width is chosen as $\sigma=0.2$, $\sigma=0.8$, $\sigma=3.2$ for the three configurations respectively; i.e.~the beam width in this section is held fixed (for each configuration) and scales proportional with system size.} To make the configurations comparable,
we choose $t_{b}=25$, $100$, and $400$, respectively. This corresponds
to roughly a doubling of the total energy contained in the plasma
system after the beam has been switched off \textcolor{black}{and an increase of the particle number by roughly 10\%}. 

In the present setup there are two parameters, the frequency $\omega$
and the corresponding amplitude $b_{1}$. The chosen figure of merit, \textcolor{black}{which we use in all numerical simulations in this work, }
is the time integrated \textcolor{black}{total} electric energy, \textcolor{black}{i.e.
\begin{equation} \text{FOM}[E] = \frac{1}{2}\int_{0}^{t_{f}}\int_0^L E^{2}(t,x)\,dx\,dt, \label{eq:fom} \end{equation} where $t_f=150$ for the small system size and $t_{f}=2t_{b}$ in the other two cases. For the small system size a time longer than $2 t_b$ is required such that the instability can fully develop (as the beam is only switched on for a relatively short duration)}. All results plotted are normalized to the case where
the beam is not modulated (i.e.,~$b_{1}=0$). Thus, we measure the
severity of the instability by how much kinetic energy is transferred
to the field. If no beam is injected into the plasma the figure of
merit is $0$ (no instability) and if no modulation is performed the
figure of merit is $1$. A smaller figure of merit thus corresponds
to a larger reduction of the \textcolor{black}{severity of the instability}.

\textcolor{black}{We note that in the sense of linear theory, an instability is primarily characterized by its growth rate. Our goal here is not necessarily to reduce this growth rate, but to (a) reduce the electric field strength at which the instability saturates and (b) to obtain a state, after the beam has been switched off, where the electric field is as small as possible (in order to facilitate heating the plasma by the available kinetic energy). The chosen figure of merits represents a compromise between the two.} 

\begin{figure*}
\begin{centering}
(a) Parameter scan for the time average of the electric energy. \textcolor{black}{The figure of merit, equation \eqref{eq:fom},} is shown for
$(\omega,b_{1})$ and $I(t)=I_{0}+b_{1}\sin(\omega t)$.
\par\end{centering}
\begin{centering}
\includegraphics[width=\textwidth]{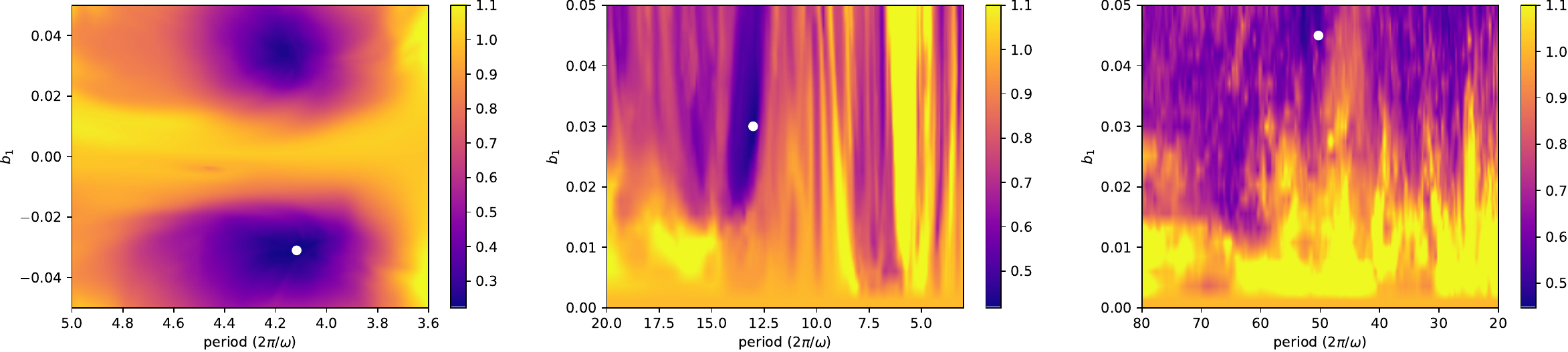}
\par\end{centering}
\begin{centering}
(b) Beam profiles (one period)
\par\end{centering}
\begin{centering}
\includegraphics[width=\textwidth]{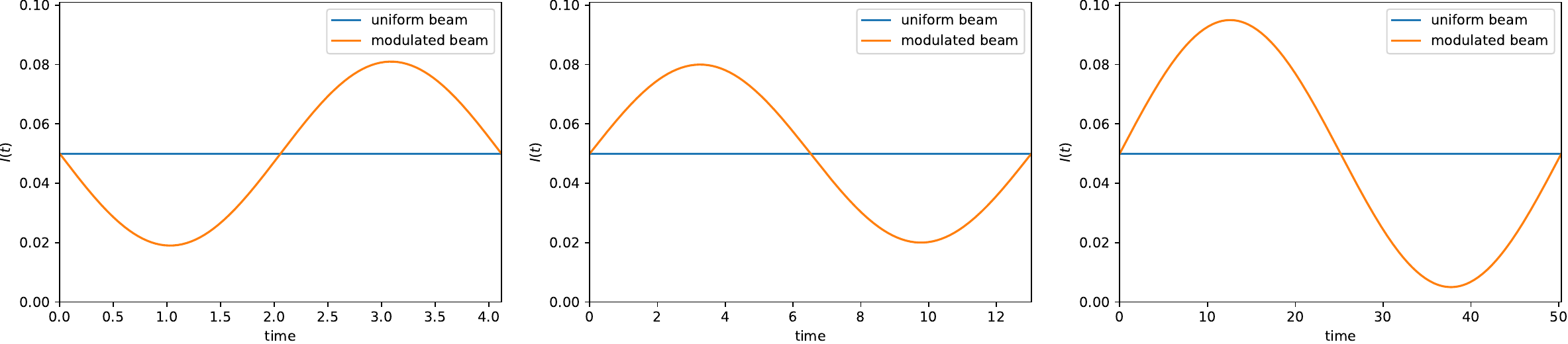}
\par\end{centering}
\begin{centering}
(c) Time evolution of the electric energy for the optimal $(\omega,b_{1})$
marked by white circles in figure (a). 
\par\end{centering}
\begin{centering}
\includegraphics[width=\textwidth]{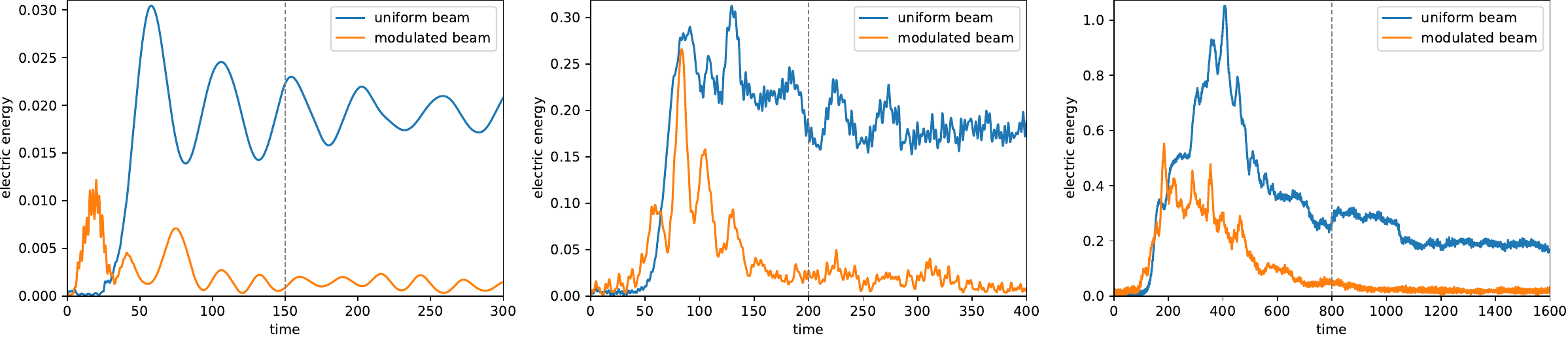}
\par\end{centering}
\caption{Numerical results for the sinusoidal beam profile. The dashed gray
line indicates the time interval $t_{f}$ over which the figure of
merit is computed. \protect\label{fig:sinusoidal}}
\end{figure*}

\begin{figure*}
\begin{centering}
\includegraphics[width=\textwidth]{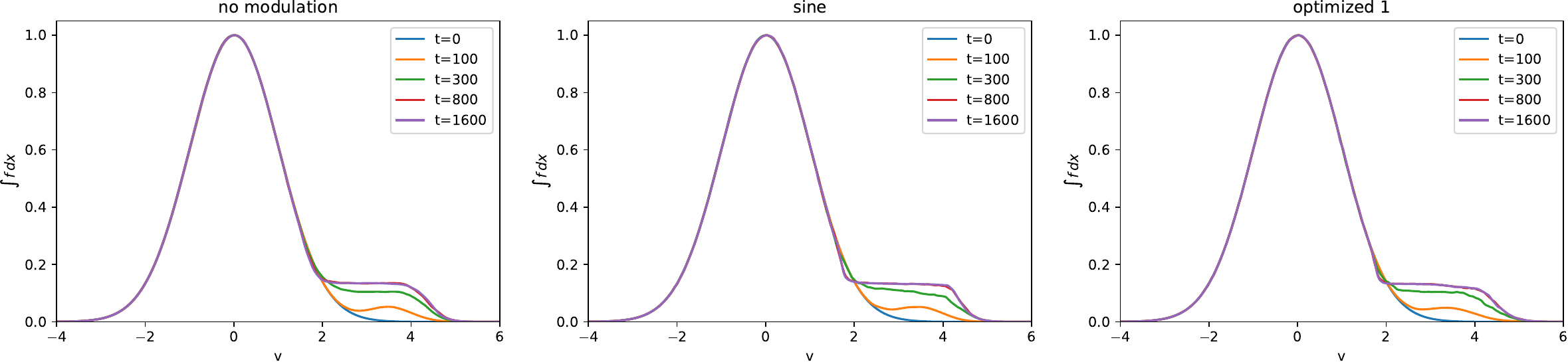}
\par\end{centering}
\caption{The spatially averaged distribution function is shown for different
times and different beam modulations for the $L=64\pi$ configuration.
For the exact specification of the beam modulation in the middle (sine)
and right (optimized 1) we refer to Table \ref{tab:optimized-coefficients}.
\protect\label{fig:evol-v}}
\end{figure*}

As can be clearly
observed from Figure \ref{fig:sinusoidal}, already a sinusoidal beam,
if an appropriate frequency and amplitude is selected, can yield a reduction in the severity of
the instability by between 56-78\%. 
In Figure \ref{fig:sinusoidal}c we consider the simulation with the
optimal frequency and amplitude for the sinusoidally modulated beam in more detail.
We observe that the reduction in the \textcolor{black}{severity of the instability} is not merely a
transient phenomenon, but persists after the beam has been switched
off. In fact, the long time behavior is excellent showing a reduction
of the electric energy by approximately 90\% compared to the unmodulated
case. This, at first glance, might seem surprising as after the beam
has been switched off the dynamics is subject to the Vlasov--Poisson
equations for which a bump-on-tail configuration is unstable. To investigate
this in more detail, we consider a plot of the velocity distribution
in Figure \ref{fig:evol-v}. What we observe is that a velocity plateau
for speeds that are roughly between two to five times the thermal
velocity develops. There is no bump and thus, as can be shown using
the Penrose stability criterion, the distribution is stable. In fact,
also in the unmodulated (i.e.~constant intensity) case the system
is driven to a state that has a velocity plateau. However, in the
unmodulated case, the primary physical mechanism is nonlinear saturation
induced by the growth of the electric field which forces
the plateau. This also results in a state with relatively large spatial
variations in the density and consequently large electric energy.
On the other hand, the modulated beam, with appropriately chosen frequency
and amplitude, drives the system into a state with a velocity plateau
that has small density variations and thus also small electric
energy.

\begin{figure*}
\begin{centering}
\includegraphics[width=\textwidth]{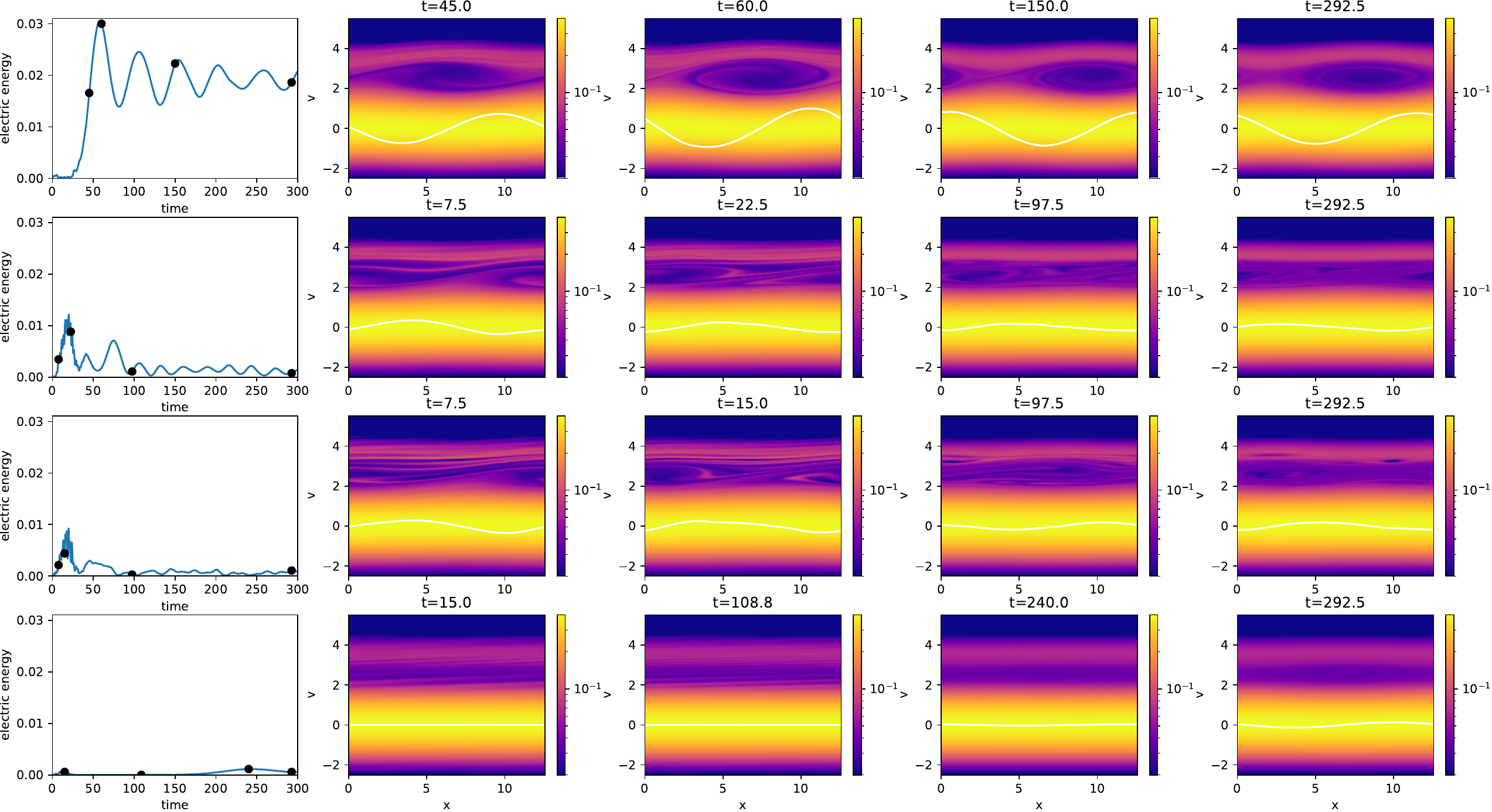}
\end{centering}
\caption{\textcolor{black}{Plots of the phase space density for the small system size ($L=4 \pi$) at different times in the evolution of the instability are shown. The times chosen are indicated with black dots in the plot of electric energy vs time on the left-hand side of the figure. From top to bottom the following beam profiles are shown: uniform, sine, optimized 1, and optimized 2. All beam profiles are plotted in Figures \ref{fig:optimization} and a detailed list of the parameters are given in Table \ref{tab:optimized-coefficients}.} \label{fig:phase-space-L1}}
\end{figure*}

\begin{figure*}
\begin{centering}
\includegraphics[width=\textwidth]{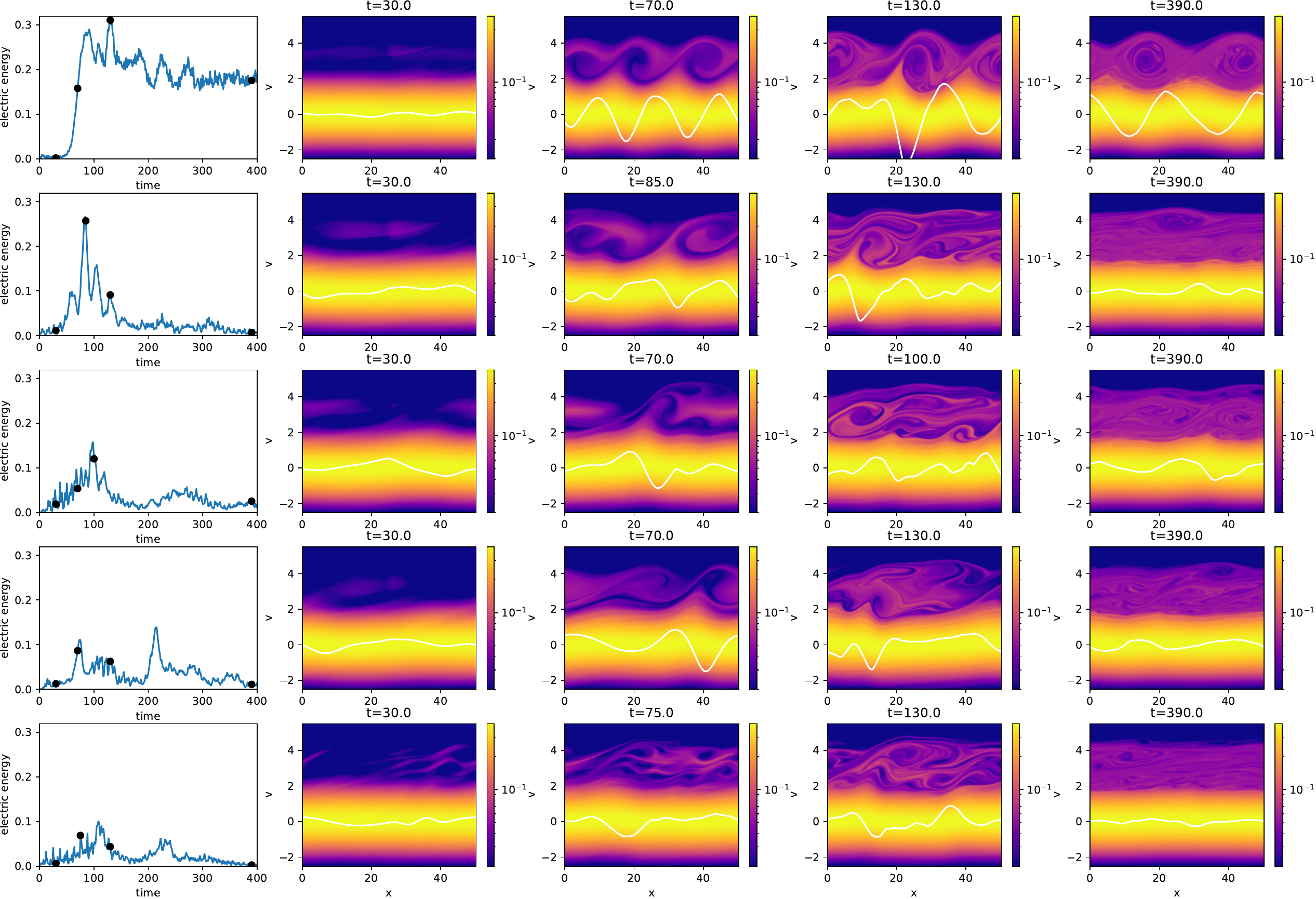}
\end{centering}
\caption{\textcolor{black}{Plots of the phase space density for the medium system size ($L=16 \pi$) at different times in the evolution of the instability are shown. The times chosen are indicated with black dots in the plot of electric energy vs time on the left-hand side of the figure. From top to bottom the following beam profiles are shown: uniform, sine, optimized 1, optimized 2, and optimized 3. All beam profiles are plotted in Figures \ref{fig:optimization} and a detailed list of the parameters are given in Table \ref{tab:optimized-coefficients}.} \label{fig:phase-space-L2}}
\end{figure*}

\begin{figure*}
\begin{centering}
\includegraphics[width=\textwidth]{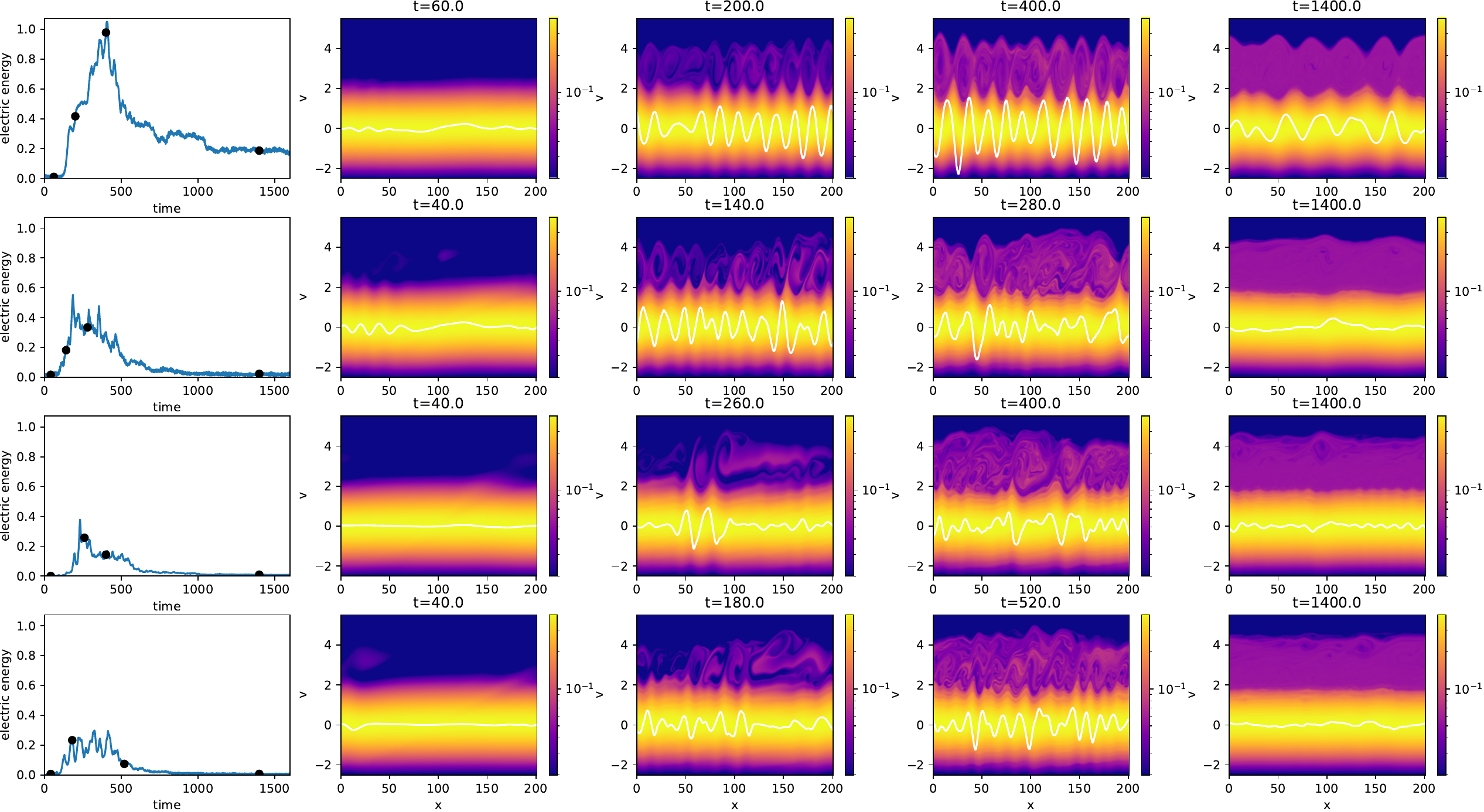}
\caption{\textcolor{black}{Plots of the phase space density for the large system size ($L=64 \pi$) at different times in the evolution of the instability are shown. The times chosen are indicated with black dots in the plot of electric energy vs time on the left-hand side of the figure. From top to bottom the following beam profiles are shown: uniform, sine, optimized 1, and optimized 2. All beam profiles are plotted in Figures \ref{fig:optimization} and a detailed list of the parameters are given in Table \ref{tab:optimized-coefficients}.} \label{fig:phase-space-L3}}
\end{centering}
\end{figure*}

\textcolor{black}{To analyze this in more detail, we have also plotted the phase space density at different points in time, see the first row for the unmodulated beam and the second row for the sinusoidal beam in Figures \ref{fig:phase-space-L1}-\ref{fig:phase-space-L3}. For the classic bump-on-tail instability the increase in the electric field (initially well described by linear theory) results in an electrostatic potential strong enough to trap particles in the wave (see, e.g., \cite{Tennyson1994}). This saturated nonlinear wave then propagates at roughly the beam velocity and can be remarkably stable over long times \cite{Balmforth2012}. Even though the beam is continuously injected into the plasma in the unmodulated case considered here, the observed behavior is very similar. The sinusoidal beam profile, on the other hand, causes different parts of the wave to merge. This starts at relatively early times in the instability and effectively breaks up the overall electrostatic structure, resulting in a spatially more homogeneous plasma.}

The important observation here is that the driving frequency
$\omega$ for the sinusoidal beam can be chosen roughly inversely
proportional to the system size. Most results that have been obtained
in the literature, on the other hand, predict a frequency close to
the plasma frequency \cite{Nakamura1970,Qin2014,Einkemmer2024b},
i.e.~independent of the system size. To understand this, let us consider
a simple phenomenological model: the Van der Pol oscillator (as used
in \cite{Nakamura1970}), i.e. a driven oscillator with a quadratic
term that models linear exponential growth for small amplitudes and
nonlinear saturation for larger amplitudes. In this case, driving
the system with the natural oscillation frequency results in suppression
of the instability. In the present setting this example is most directly
related to the smallest domain size ($L=4\pi$). In this case there
is only a single unstable mode. Solving the dispersion relation for
the bump-on-tail instability (which obviously neglect time dependent
beam propagation effects) gives $\omega_{\text{osc}}\approx1.45$
which matches well with $\omega\approx1.53$ as given in Figure \ref{fig:sinusoidal}.
Non phenomenological models can be derived. However, they usually
rely on decoupling the spatial modes (as Fourier techniques are used).
This misses an important effect. Namely, that to excite oscillations
with the largest possible wavelength in space, a frequency of $\omega=v_{b}/L$
is sufficient. In fact, if $\omega$ is not an exact multiple of $v_{b}/L$
multiple modes are excited. This allows for manipulation of the bump-on-tail
feature with frequencies that are large compared to the plasma frequency,
leading to the observed results. This is immensely useful as it means
that for large systems we do not need to modulate the beam at frequencies
comparable to the (usually very high) plasma frequency.

\section{Results for general beam profiles\protect\label{sec:general-beams}}

We now turn our attention to the case of more general beam profiles.
That is, our goal is to find $I(t)$, $\sigma$, and $v_{th,b}$ (subject
to the constraints outlined in section \ref{sec:model}), that reduce
the figure of merit as much as possible. This can be formulated as
an optimization problem
\begin{equation}
\min_{I(t),\sigma,v_{th,b}}\int_{0}^{t_{f}}\int_0^L \left(E[f_{I,\sigma,v_{th,b}}](t)\right)^{2}\,dx\,dt,\label{eq:optimization-problem}
\end{equation}
where $f_{I,\sigma,v_{th,b}}$ denotes the solution of equation (\ref{eq:vlasov-with-source})
for the specified parameters. The beam intensity is subject to the
constraint that $I(t)\geq0$ and a fixed total intensity $\int_{0}^{t_{b}}I(t)\,dt=I_{0}t_{b}$.
However, for such a scheme to be useful in practice, we have to constrain $I(t)$ such that it varies in a reasonable manner. There are
many possibilities to do this. We enforce this here by only allowing
a couple of Fourier modes. That is, we parameterize $I(t)$ as
\[
I(t)=\text{max}\left(I_{0}+\sum_{k=1}^{K-1}a_{k}\cos(k\omega_{0}t)+\sum_{k=1}^{K}b_{k}\sin(k\omega_{0}t),0\right).
\]
Taking the maximum is necessary because some combination of parameters
$(a_{k},b_{k})$ can result in negative beam intensities, which is
clearly unphysical. By taking the maximum with $0$, we violate the
constraint of fixed total intensity. However, we always have $\int_{0}^{t_{b}}I(t)\,dt>I_{0}t_{b}$
and thus finding such a result in the optimization would indicate
that a smaller electric energy with a (slightly) higher beam intensity
has been achieved, which is clearly a favorable outcome. 

We emphasize that, as Figure \ref{fig:sinusoidal}a already makes
clear, the landscape of the optimization problem is quite rough. Thus,
we have a global optimization problem with many local minima and a
total of $2K+2$ parameters. In the following we will present the
obtained results. However, some more details on our approach to solve
this optimization problem in an efficient way are outlined in section
\ref{sec:methods}.

\begin{figure*}
\begin{centering}
(a) Beam profiles (one period of the beam profile with the lowest frequency)
\par\end{centering}
\begin{centering}
\includegraphics[width=\textwidth]{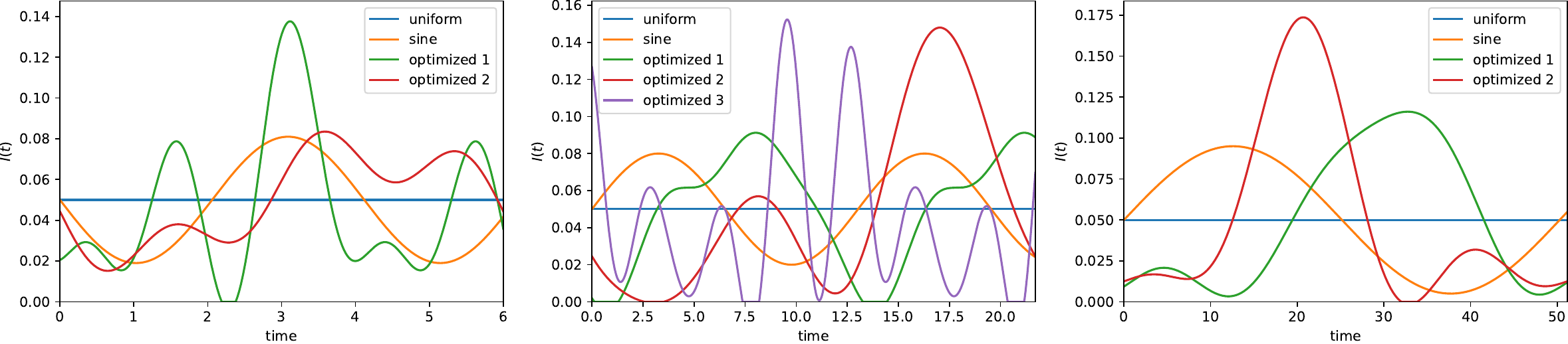}
\par\end{centering}
\begin{centering}
(b) Time evolution of the electric energy
\par\end{centering}
\begin{centering}
\includegraphics[width=\textwidth]{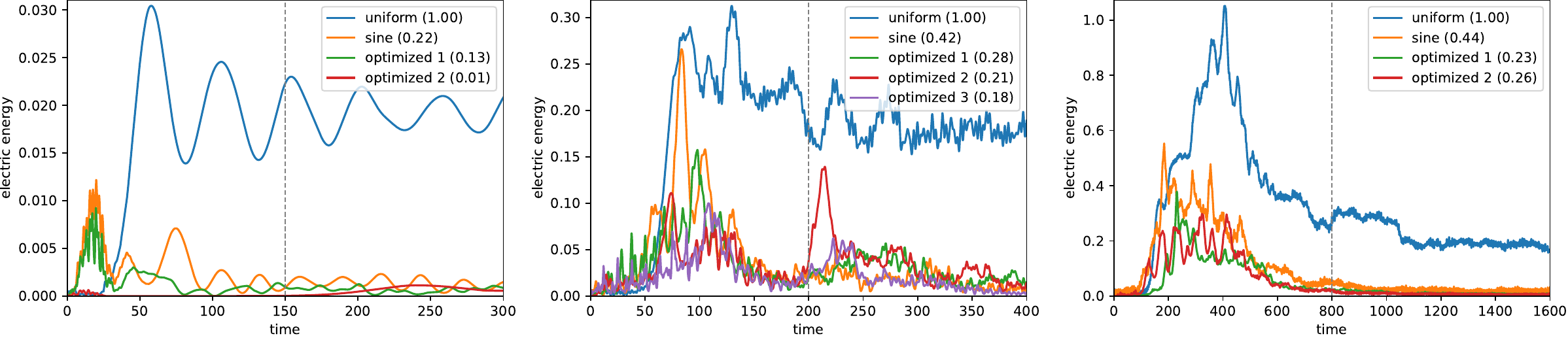}
\par\end{centering}
\caption{Comparison of the beam profiles found by numerical optimization (denoted
by optimized 1, optimized 2, and optimized 3; the parameters can be
found in Table \ref{tab:optimized-coefficients}) with the optimal
sinusoidal beam profiles from section \ref{sec:sinusoidal}. The figure
of merit is shown in parentheses in the legend and the dashed gray
line indicates the time interval $t_{f}$ over which the figure of
merit is computed. \protect\label{fig:optimization}}
\end{figure*}

In Figure \ref{fig:optimization} we present the best candidates that
have been obtained from the numerical optimization and compare them
to the sinusoidal beam profile found in the previous section. Note
that the global optimization algorithm does not give a unique solution
but a number of candidates. We have selected the best (according to
the figure of merit) candidates where \emph{optimized 1} only optimizes
for the beam profile (but keeps $\sigma$ fixed) and
\emph{optimized 2} simultaneously optimizes the beam profile as well
as $\sigma$ and $v_{th,b}$. For the intermediate problem size we
have also included an optimization run (called \emph{optimized 3}
in the plot) with $K=5$ (in all other configuration $K=3$ is used).
The parameters of all configurations can be found in Table \ref{tab:optimized-coefficients}.
For the two larger system sizes we roughly observe a reduction of
the severity of the instability by approximately 75-80\%, which is a
further reduction compared to the sinusoidal beam profile by approximately
a factor of $2$. The outlier here is the smallest system size, where
the figure of merit decreases by more than 99\%. We also see excellent
behavior with respect to long time behavior (i.e.~after the beam
has been switched off). The frequency required to achieve this is,
again, roughly inversely proportional to the system size.

\textcolor{black}{We also note that the beam width $\sigma$ has a relatively minor effect on the result. We have restricted the optimization such that $\sigma$ is at least $0.2$, $0.8$, and $3.2$ for the small, medium, and large system sizes, respectively. While the optimal solution obtained generally has a value close to this, we have included a candidate with $\sigma\approx 6.26$ (\emph{optimized~2} for the large problem size) to demonstrate that even for larger beam widths the achieved figure of merit only deteriorates slightly.}

We further study the velocity dependence of a optimized solution in
Figure \ref{fig:evol-v} on the right. As expected the solution is
driven to a state with a plateau in the velocity distribution. In
fact, we observe that the dynamics produces almost no bump-on-tail
feature in this case. \textcolor{black}{This can be seen in more detail by the phase space plots in Figures \ref{fig:phase-space-L1}-\ref{fig:phase-space-L3}, where we observe that the beam profiles found by optimization are even more effective in merging the parts of the electrostatic waves and thus reducing the amplitude of the instability.}

\section{Robustness of the beam profiles to perturbation}

\begin{figure*}
\begin{centering}
\includegraphics[width=0.33\textwidth]{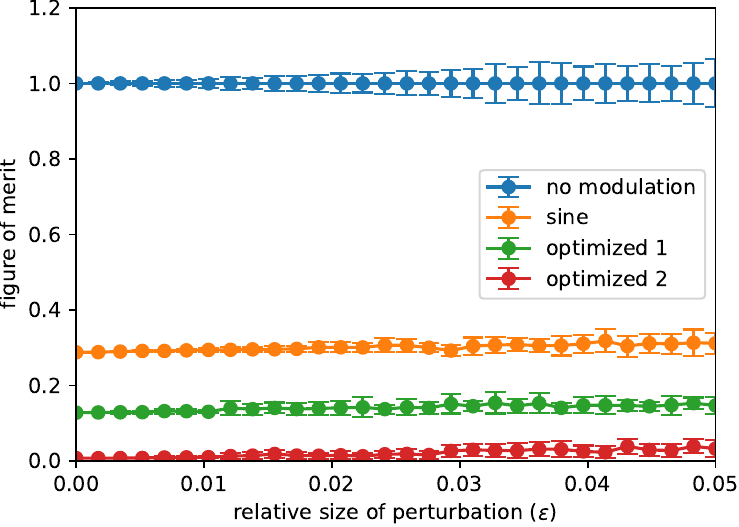}\includegraphics[width=0.33\textwidth]{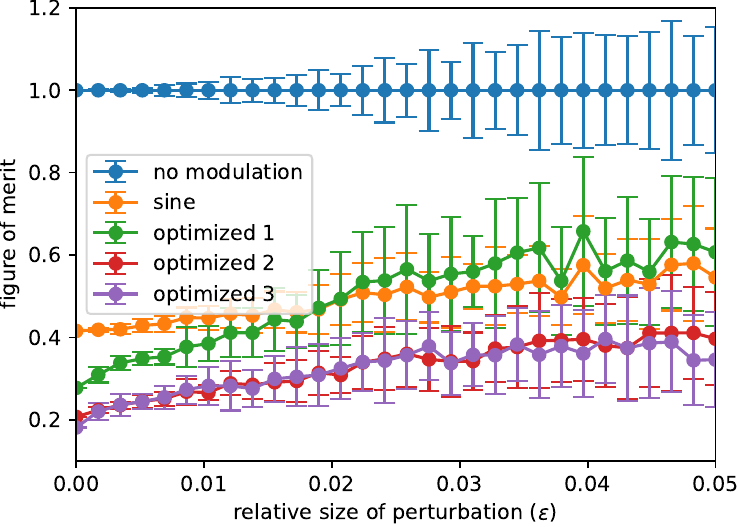}\includegraphics[width=0.33\textwidth]{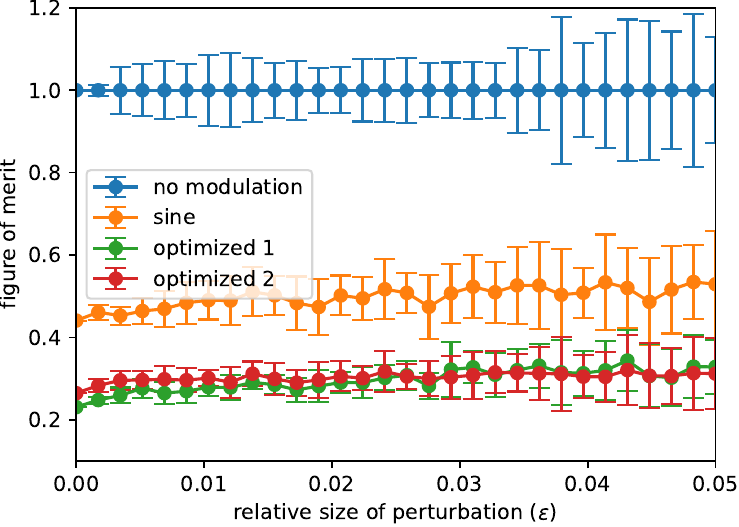}
\par\end{centering}
\caption{The figure of merit as a function of the relative size of the perturbation
$\epsilon$ in the parameters that determine the beam profile is shown.
For each $\epsilon$ we consider $20$ random perturbations and their
average figure of merit as well as the standard deviation is shown.
Since the perturbation also changes the figure of merit for the case
with an unmodulated beam, we have normalized all plots to the corresponding
average value for the unmodulated beam at that size of the perturbation.\protect\label{fig:robust-coeff}}
\end{figure*}

A control scheme would not be very useful in practice if even minor
perturbations, e.g.~due to imperfection in the control of the beam,
would be sufficient to negate its effect. This is what we will study
in this section. We run the simulations as before, but with the parameters
in Table \ref{tab:optimized-coefficients} perturbed by multiplying
them with uniformly distributed random numbers in the range $[1-\epsilon,1+\epsilon]$.
This gives a modified beam profile and the corresponding results are
shown in Figure \ref{fig:robust-coeff}. We observe that the figure
of merit is remarkably robust at least for perturbations of up to
5\% in all parameters. The only outlier is the optimized 1 beam profile
for the intermediate system size, which relatively quickly deteriorates
to the level of the sinusoidal beam profile. Thus, different beam
profiles can behave differently with respect to such perturbations.
In principle, one could also perform the optimization taking this
into account. For example, by penalizing beam profiles that are not
robust. However, with one exception all beam profiles found using
the optimization algorithm (that only takes the figure of merit into
account) turned out to be robust.

\section{Methodology\protect\label{sec:methods}}

Finding the parameters $(\omega_{0},a_{k},b_{k},\sigma,v_{th,b})$
of the optimization problem given in equation (\ref{eq:optimization-problem})
requires a global optimization algorithm in order to avoid being stuck
in local minima. We use a modification of the genetic optimization
algorithm found in\texttt{ }\texttt{scipy}\texttt{.optimize.differential\_evolution}
\cite{scipy}. The genetic optimization algorithm can be setup such
that in each generation the figure of merit (or fitness) for all candidate
solutions (which are sets of parameters that determine the beam profiles)
are evaluated at the same time. This is beneficial from a computational
point of view, as calling individual runs of a one-dimensional Vlasov--Poisson
solver has a relatively large overhead due to the small problem size.
In order to avoid this we have developed a C++ code that solves equation
(\ref{eq:vlasov-with-source}) in a batched manner (i.e.~multiple
beam profiles at the same time). The code uses the Kokkos performance
portability programming ecosystem \cite{kokkos} and can thus make
use of modern computing systems equipped with graphic processing units
(GPUs).

In the optimization algorithm the simulations are run with 512 grid
points in the spatial direction, 256 grid points in the velocity direction,
and a time step size $\Delta t=0.02$. A splitting based semi-Lagrangian
scheme with 9th order Lagrange interpolation is employed. However,
in order to make sure that all results obtained from the optimization
algorithm are independent of the numerical parameters used, all the
data required for the plots in this paper have been run with increased
resolution (1024 grid points in the spatial, 512 grid points in the
velocity direction, and a time step size $\Delta t=0.01$). 

\section{Discussion}

In this work, we have investigated how much control can be achieved
by modulating the intensity of a beam in an electrostatic beam-plasma
instability. We have found beam profiles that drastically reduce the
severity of the instability. In all configurations studied here, control
schemes have been obtained that reduce the time averaged electric
energy (a measure of the severity of the instability) by at least
75\% compared to its original value.

A pertinent question is whether such or a related control scheme can
be applied in practice. In that regard, it is encouraging that, as
we have demonstrated, a significant reduction in the severity of the
instability can be observed with a beam modulation that
\begin{itemize}
\item requires no feedback (i.e.~measurements) during the evolution of
the instability. Control without feedback is possible since the modulation
of the beam (i.e.~how the system is driven) has a significant influence
on the velocity distribution. In fact, the bump-on-tail feature can
largely be avoided for the optimized beam profiles and even for sinusoidally modulated beams is significantly
reduced;
\item the frequency at which we need to modulate the beam scales favorably
with system size. This is important in practice as the frequency of
the beam determines how fast the control (here the beam intensity)
has to be actuated and consequently how easy it is to implement such
a scheme. This is, in particular, favorable as in many systems of
interest where beam heating is employed the size of a plasma system
can be many thousands of Debye lengths;
\item is able to drive the system to a state that remain stable after the
beam has been switched off. This state has a plateau in velocity space
and almost uniform density in space, which implies that little kinetic
energy from the beam is transferred to the field;
\item is robust with respect to perturbations in the beam profile. That
is, imperfections in the beam profile (as are unavoidable in practice)
do not significantly diminish the performance of the control scheme.
\item \textcolor{black}{does not require a tight focusing of the beam. In fact, we found that the beam width (for a beam width of up to 3\% of the domain size) has only a small influence on the effectiveness of the control.}
\end{itemize}
We emphasize that our control scheme relies on the time evolution
of the bump-on-tail feature to be effective. This is different form
the situation in, e.g.~\cite{Qin2014}, where two streams are already
established and are then subsequently modulated by applying an external
electric field. We take the viewpoint that, at least in beam heating
applications, assuming an already established bump-on-tail feature
is more a modeling artifact often required to make theoretical progress
than a reasonable experimental setup.

Let us also point out that we do not model the actual heating (i.e.~the
thermalization of the velocity distribution by collisions) here. In
the spatially homogeneous case, it can be easily inferred that the
system relaxes to a Maxwellian with increased temperature after the
beam has been switched off. The numerical simulation show
that a small degree of spatial variation remains. Nevertheless, since
the state obtained after the beam has been switched off is stable,
we expect the same result to hold true. We consider this as future
work.

For the present study we have used a simple one-dimensional model
with stationary ions. In many applications of practical relevance
more faithful models are required that are much more complicated and
more challenging numerically (e.g. due to the timescale separation
between ions and electrons, up to six dimensions in phase space, the
beam consisting of both electrons and ions, etc.). This presents a
significant challenge as in order to solve the global optimization
problem many simulations have to be conducted. Thus, in order to
find good beam profiles in such a situation would most likely require
either some type of complexity reduction techniques/reduced models
or large-scale supercomputers. We also note that in more realistic
problems there are many more parameters that could be part of the
optimization algorithm (e.g.~the angle at which a beam penetrates
into the plasma), which conceivable could help in \textcolor{black}{reducing the severity of the instability}.

\section*{Acknowledgements} \textcolor{black}{We would like to thank the anonymous reviewers for their insightful comments that helped to improve the paper. We would also like to thank Qin Li (University of Wisconsin, Madison), Li Wang (University of Minnesota) and Yunan Yang (Cornell University) for fruitful discussions on optimization algorithms and plasma stability.}

\begin{table*}
\footnotesize \begin{tabular}{r|c|c|c}
 & $L=4 \pi$ & $L=16 \pi$ & $L=64 \pi$ \\
\hline
sine & \begin{tabular}{ll}$\omega_{0}=1.5258$ & \\ $\sigma=0.2$ & $v_{\mathrm{th},b}=0.5$ \\ $I_0=0.05$ & $b_1=-0.031$ \end{tabular} & \begin{tabular}{ll}$\omega_{0}=0.48221$ & \\ $\sigma=0.8$ & $v_{\mathrm{th},b}=0.5$ \\ $I_0=0.05$ & $b_1=0.03$ \end{tabular} & \begin{tabular}{ll}$\omega_{0}=0.12491$ & \\ $\sigma=3.2$ & $v_{\mathrm{th},b}=0.5$ \\ $I_0=0.05$ & $b_1=0.045$ \end{tabular}\\
\hline
optimized 1 & \begin{tabular}{ll}$\omega_{0}=1.5515$ & \\ $\sigma=0.2$ & $v_{\mathrm{th},b}=0.5$ \\ $I_0=0.05$ & $b_1=-0.026653$\\ $a_1=-0.00091327$ & $b_2=-0.02506$\\ $a_2=-0.028601$ & $b_3=0.029389$ \end{tabular} & \begin{tabular}{ll}$\omega_{0}=0.47789$ & \\ $\sigma=0.8$ & $v_{\mathrm{th},b}=0.5$ \\ $I_0=0.05$ & $b_1=-0.015525$\\ $a_1=-0.037118$ & $b_2=-0.00095985$\\ $a_2=-0.010956$ & $b_3=-0.0065273$ \end{tabular} & \begin{tabular}{ll}$\omega_{0}=0.12455$ & \\ $\sigma=3.2$ & $v_{\mathrm{th},b}=0.59837$ \\ $I_0=0.05$ & $b_1=-0.035227$\\ $a_1=-0.042398$ & $b_2=0.016937$\\ $a_2=0.0017211$ & $b_3=0.0094516$ \end{tabular}\\
\hline
optimized 2 & \begin{tabular}{ll}$\omega_{0}=1.0472$ & \\ $\sigma=1$ & $v_{\mathrm{th},b}=0.6$ \\ $I_0=0.05$ & $b_1=-0.024724$\\ $a_1=-0.007559$ & $b_2=-0.001239$\\ $a_2=0.0019077$ & $b_3=-0.014125$ \end{tabular} & \begin{tabular}{ll}$\omega_{0}=0.28944$ & \\ $\sigma=0.8$ & $v_{\mathrm{th},b}=0.6$ \\ $I_0=0.05$ & $b_1=-0.045485$\\ $a_1=0.0046448$ & $b_2=-0.031921$\\ $a_2=-0.030133$ & $b_3=0.014797$ \end{tabular} & \begin{tabular}{ll}$\omega_{0}=0.12278$ & \\ $\sigma=6.262$ & $v_{\mathrm{th},b}=0.59782$ \\ $I_0=0.05$ & $b_1=0.035498$\\ $a_1=-0.047241$ & $b_2=-0.042762$\\ $a_2=0.0097479$ & $b_3=0.021815$ \end{tabular}\\
\hline
optimized 3 &  & \begin{tabular}{ll}$\omega_{0}=0.48407$ & \\ $\sigma=0.80088$ & $v_{\mathrm{th},b}=0.59999$ \\ $I_0=0.05$ & $b_1=-0.020504$\\ $a_1=0.023424$ & $b_2=-0.00080311$\\ $a_2=-0.0064016$ & $b_3=0.011956$\\ $a_3=0.014704$ & $b_4=-0.022388$\\ $a_4=0.044865$ & $b_5=-0.013898$ \end{tabular} & \\
\hline \end{tabular}

\caption{The parameters of the optimized beam profiles that are discussed in
sections \ref{sec:sinusoidal} and \ref{sec:general-beams} are listed.\protect\label{tab:optimized-coefficients} }
\end{table*}

\bibliography{plasma-physics,literature}

\end{document}